\documentclass[graybox]{svmult}

\usepackage{mathptmx}       
\usepackage{helvet}         
\usepackage{courier}        
\usepackage{type1cm}        
\usepackage{makeidx}         
\usepackage{graphicx}        
\usepackage{multicol}        
\usepackage[bottom]{footmisc}
\usepackage{cite}

\makeindex             

\begin{document}

\title*{Vector meson spectral function and dilepton rate in an effective mean field model}

\author{Chowdhury Aminul Islam, Sarbani Majumder, Najmul Haque, Munshi G. Mustafa}

\institute{Chowdhury Aminul Islam, Sarbani Majumder, Munshi G. Mustafa \at Saha Institute of Nuclear Physics, 1/AF, Bidhannagar, Kolkata-700064, India, \email{chowdhury.aminulislam@saha.ac.in, sarbani.majumder@saha.ac.in, munshigolam.mustafa@saha.ac.in}
\and Najmul Haque \at Department of Physics, Kent State University, Kent, OH 44242, United States, \email{nhaque@kent.edu}}

\maketitle

\abstract*{We have studied the vector meson spectral function (VMSF) in a hot and dense medium within an effective QCD model namely the Nambu$\textendash$Jona-Lasinio (NJL) and its Polyakov Loop extended version (PNJL) with and without the effect of isoscalar vector interaction (IVI). The effect of the IVI has been taken into account using the ring approximation. We obtained the dilepton production rate (DPR) using the VMSF and observed that at moderate temperature it is enhanced in the PNJL model as compared to the NJL and Born rate due to the suppression of color degrees of freedom.}

\abstract{We have studied the vector meson spectral function (VMSF) in a hot and dense medium within an effective QCD model namely the Nambu$\textendash$Jona-Lasinio (NJL) and its Polyakov Loop extended version (PNJL) with and without the effect of isoscalar vector interaction (IVI). The effect of the IVI has been taken into account using the ring approximation. We obtained the dilepton production rate (DPR) using the VMSF and observed that at moderate temperature it is enhanced in the PNJL model as compared to the NJL and Born rate due to the suppression of color degrees of freedom.}

\section{Introduction}
\label{sec:1}
Under extreme conditions, {\textit {i.e.}}, at high temperature and/or high density Quantum Chromodynamics (QCD) exhibits a very rich and interesting phase structure. It is believed that at these extreme conditions normal hadronic matter transforms into a deconfined state of strongly correlated quark gluon plasma (QGP) \cite{Muller:1983ed,Heinz:2000bk} . A lot of efforts have been put to create and explore this noble state of matter in laboratory through experiments of heavy ion collisions such as Relativistic Heavy Ion Collider (RHIC) at Brookhaven National Lab (BNL) \cite{Adams:2005dq,Adcox:2004mh} and the Large Hadron Collider (LHC) at the European Organization for Nuclear Research (CERN) \cite{Carminati:2004fp,Alessandro:2006yt}. There is also an upcoming fixed target experiment named as Facility for Antiproton and Ion Research (FAIR) at the Gesellschaft f\"ur Schwerionenforschung (GSI)\cite{Friman:2011zz}. Various measurements at RHIC BNL \cite{Adare:2006ti} and LHC CERN \cite{Aamodt:2010pa} have indicated 
a 
strong hint of the creation of a strongly correlated QGP \cite{Islam:2014tea}.

Many properties of the deconfined, strongly interacting matter are reflected in the structure of the correlation function (CF) and its spectral representation \cite{Forster(book):1975HFBSCF}. The temporal part of the CF reflects response of the conserved density fluctuations whereas the spatial part reveals the information on the masses and width. We can construct mesonic current-current CF's from meson currents and they can be of different types such as scalar, pseudoscalar, vector and pseudovector \cite{Davidson:1995fq}. The spectral representation of vector meson current-current CF is related to the differential lepton pair production rate \cite{Kapusta_Gale(book):1996FTFTPA}. These leptons, being weakly interacting, carry informations about the earlier stage of the produced QGP. Keeping this in mind we construct the vector meson spectral function (VMSF) and obtain the dilepton production rate (DPR) from it.

Now there are many theoretical tools which can be used to study these CF's and their spectral representations  - as for example first principle calculation like Lattice QCD (LQCD) \cite{Ding:2010ga}, perturbative calculation known as perturbative QCD (PQCD) \cite{Braaten:1990wp,Greiner:2010zg,Mustafa:1999dt} and effective QCD models \cite{Klevansky:1992qe,Buballa:2003qv}. These theoretical tools have their own merits and demerits. This present work is based on the effective QCD model namely the Nambu$\textendash$Jona-Lasinio (NJL) and its Polyakov Loop extended version (PNJL). We have studied these models with and without the effects of isoscalar vector interaction (IVI). The effect of IVI has been taken into account through ring approximation. In Sec. \ref{sec:2} we write down the equation to calculate the differential dilepton rate from VMSF. The results are discussed in Sec. \ref{sec:3} and in Sec. \ref{sec:4} we conclude.

\section{Theoretical Framework}
\label{sec:2}
 The differential DPR is related to the VMSF , $\sigma_V$, as \cite{Karsch:2000gi} 

\begin{equation}
  \frac{dR}{d^4xd^4Q}  =\frac{5\alpha^2} {54\pi^2} \frac{1}{M^2} 
\ \frac{1}{e^{\omega/T}-1} \ \sigma_V(\omega, {{\vec q}}) \ , \label{l_eq.redilep_spec}
\end{equation}

where $\alpha$ is the fine structure constant, $Q \equiv (q_{0}=\omega,\vec{q})$ is the external four momentum and  $M=\sqrt{\omega^2-q^2}$ is the invariant mass of the lepton pair.

\section{Results}
\label{sec:3}

\begin{figure}[h]
\begin{center}
\includegraphics[scale=0.5]{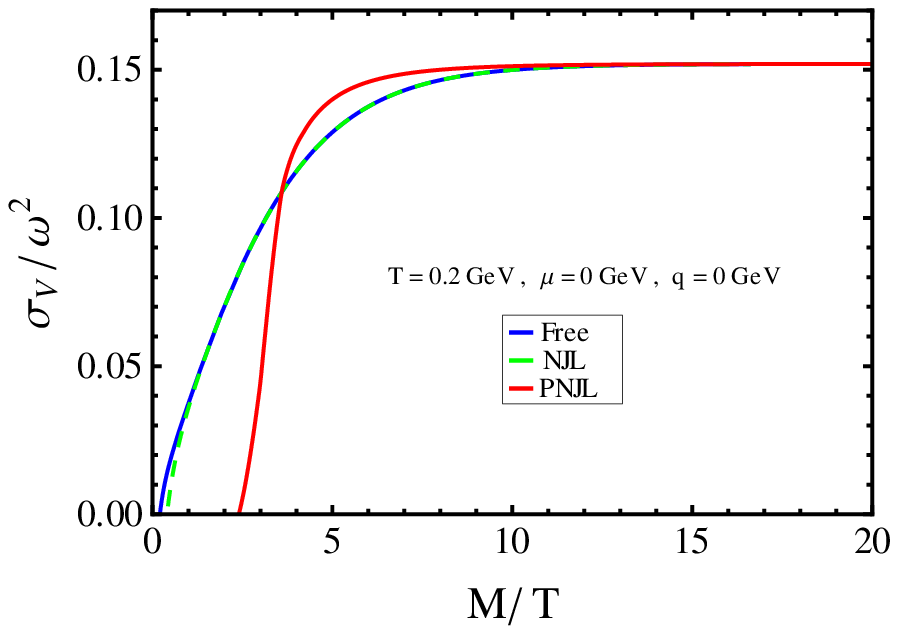}
\includegraphics[scale=0.5]{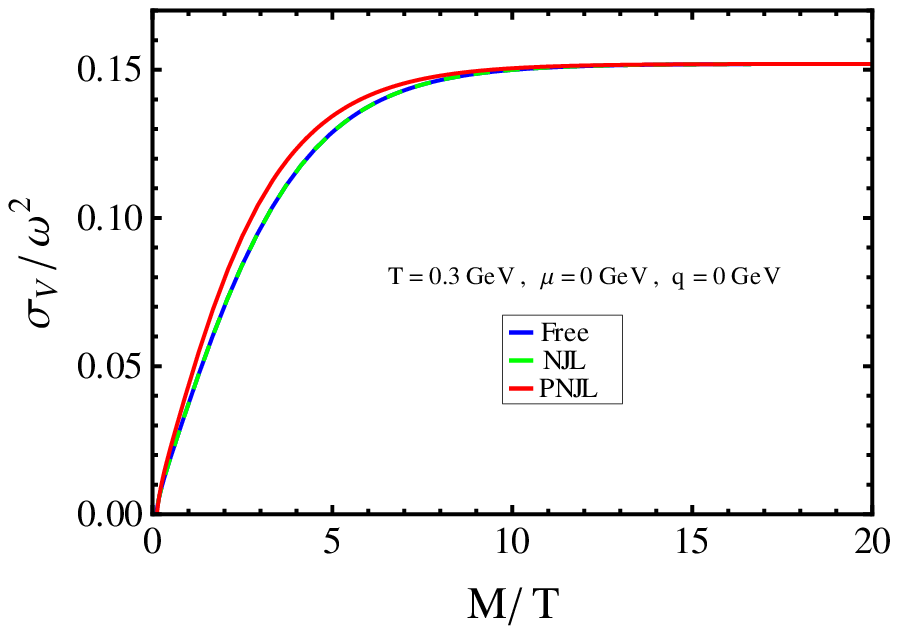}
\caption{Scaled vector spectral function $\sigma_V/\omega^2$ as a function of scaled 
invariant mass, $M/T$ with $q= \mu =0$ and $G_V/G_S=0$ for $T=200$ MeV (left panel) and $T=300$ MeV (right panel).}
\label{fg.spect_gv0_q0_mu0}  
\end{center}
\end{figure}

Here we briefly discuss our findings. The details can be found in Ref.\cite{Islam:2014sea}. In Fig. \ref{fg.spect_gv0_q0_mu0} the scaled spectral functions (SF) without the IVI are shown. At $T = 200$ MeV we observe that the PNJL SF has a larger threshold than the NJL one because of the larger quark mass. This effective quark mass decreases as we increase the temperature and that is why at $T=300$ MeV the thresholds for NJL and PNJL are almost the same. We also observe that for the PNJL case the SF has been enhanced. This can be understood with the following argument. For zero external momentum ($q$) and zero chemical potential ($\mu$) the SF is proportional to $[1-2f(E_p)]$, where $f(E_p)$ is the fermion distribution function. Now the presence of the Polyakov Loop (PL) fields suppress the color degrees of freedom, or in other words suppress $f(E_p)$ and thus enhances the SF.

\begin{figure}[h]
\sidecaption[t]
\includegraphics[scale=0.5]{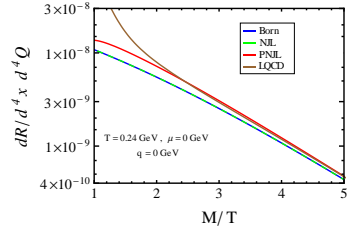}
\caption{Comparison of  dilepton rates  as a function of scaled invariant mass $M/T$ for $T=240$ MeV with external momentum 
$q=0$, quark chemical potential $\mu=0$ and $G_V/G_S=0$. The LQCD rate is from Ref.\cite{Ding:2010ga}}
\label{fg.dilep_gv0_q0_mu0}    
\end{figure}

Once we have the SF, the dilepton rate can be calculated from it using Eq. \ref{l_eq.redilep_spec}. The differential dilepton rates are shown in Fig. \ref{fg.dilep_gv0_q0_mu0}. We notice that the dilepton rate for PNJL model is increased as compared to the NJL and Born rate. We have further compared our findings with the available quenched LQCD results\cite{Ding:2010ga}.

\begin{figure}[h]
\begin{center}
\includegraphics[scale=0.5]{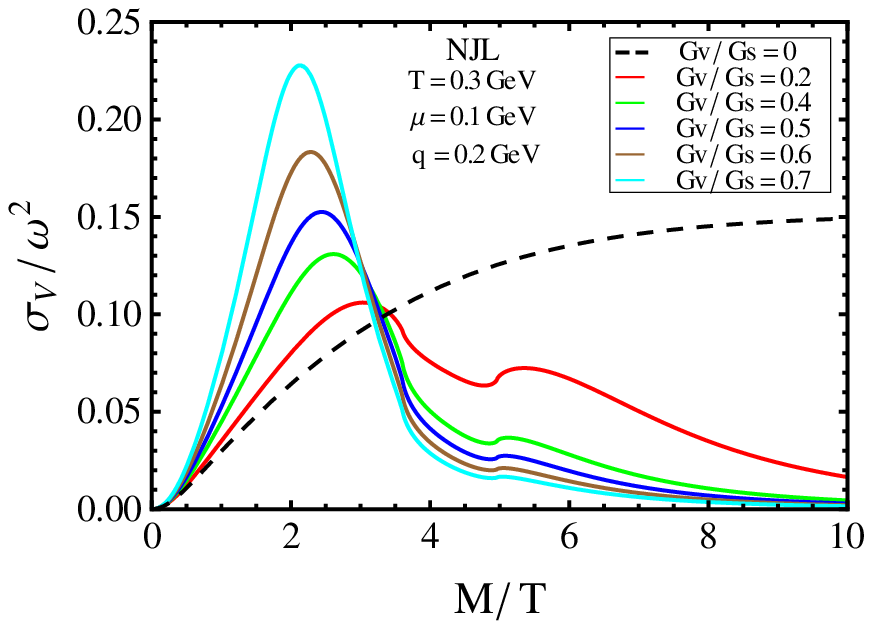}
\includegraphics[scale=0.5]{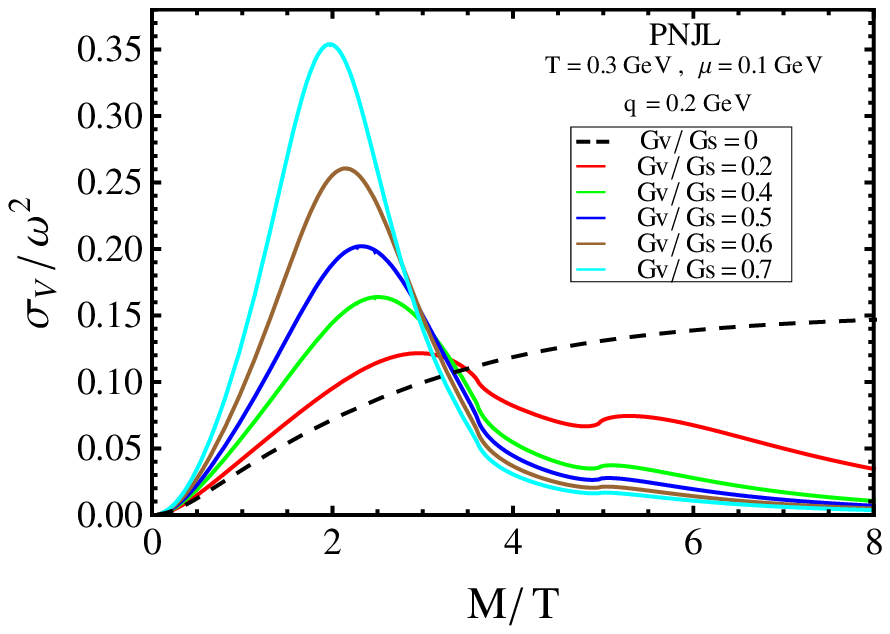}
\caption{Scaled spectral function  as a function of  $M/T$ 
for $T=300$ MeV, $\mu=100$ MeV,  $q=200$ MeV and a set of values of $G_V/G_S$; NJL (left panel) and PNJL (right panel).}
\label{fg.spect_t3_q2_mu1}  
\end{center}
\end{figure}

The SF's with the inclusion of IVI are shown in Fig. \ref{fg.spect_t3_q2_mu1}. It is clear that in the presence of IVI, bound states start to form. The spectral strength increases as we increase the strength of the IVI\footnote{ The strength of the IVI is represented by $G_V$ in terms of $G_S$, where $G_S$ and $G_V$ denote coupling constants of the scalar and vector type four-quark interactions respectively.} and for a given $G_V$ the spectral strength for the PNJL is greater than NJL, which is again attributed to the presence of PL fields.

\begin{figure}[h]
\begin{center}
\includegraphics[scale=0.5]{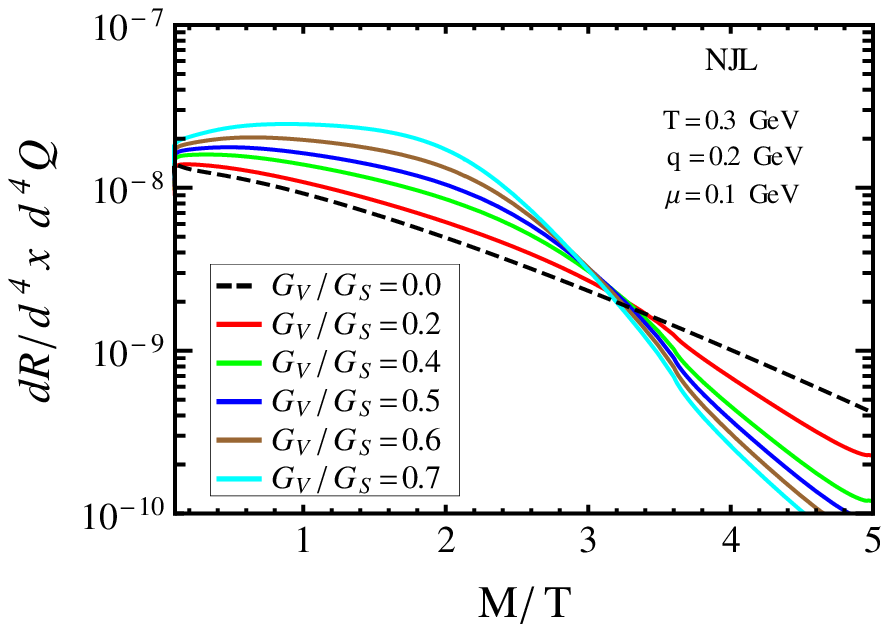}
\includegraphics[scale=0.5]{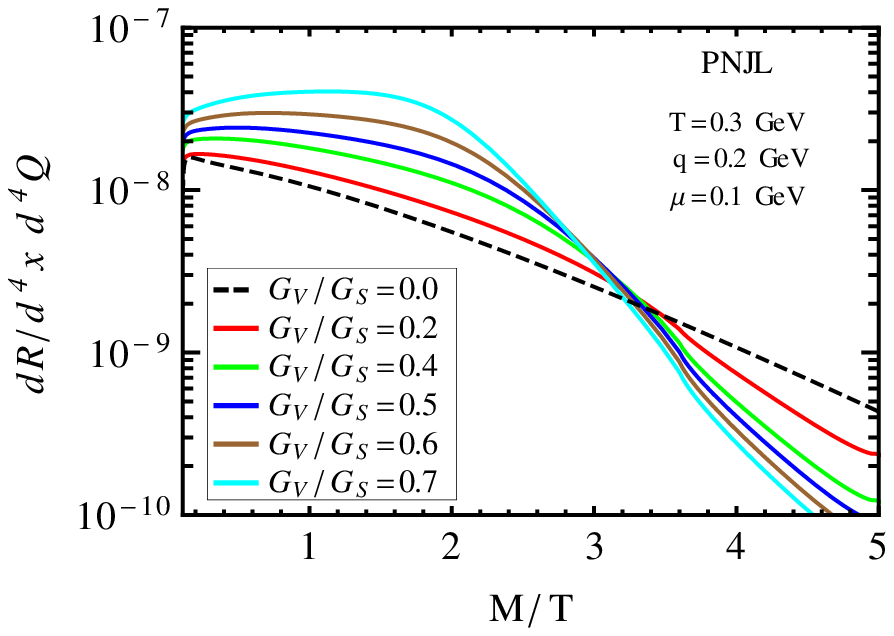}
\caption{Dilepton rates  as a function of  $M/T$ for a set of values of $G_V/G_S$  with $T=300$ MeV, $q=200$ MeV and $\mu=100$ MeV; NJL (left panel) and PNJL (right panel).}
\label{fg.dilep_gv_t3}  
\end{center}
\end{figure}

In Fig. \ref{fg.dilep_gv_t3} the differential dilepton rates are shown for both NJL and PNJL models. The dilepton rates increase as the strength of the IVI increases which was expected if we look back at the corresponding plots of spectral function in Fig.   \ref{fg.spect_t3_q2_mu1}.

\section{Summary} 
\label{sec:4}
We have investigated the behavior of the VMSF and its spectral representation in a hot and dense medium. The differential dilepton rate has been calculated from the SF for both with and without the IVI. We observed that without the IVI the SF's in NJL become quantitatively equal to those of free field theory. For PNJL they are enhanced as compared to both the NJL and the free field theory. This enhancement is reflected in the corresponding plots of dilepton rates. Suppression of the color degrees of freedom due to the presence of PL fields is the reason behind this enhancement. This suggests that some nontrivial correlation exists among the color charges in the deconfined phase. Thus it is expected that the dilepton rate in a strongly coupled QGP is more than that in a weakly coupled one. We have compared some of our results with the available quenched LQCD data and the other results can be tested in future when LQCD includes the dynamical fermions.

\begin{acknowledgement}
CAI was supported financially by the University Grants Commission, India; SM, NH and MGM were supported by the Department of Atomic Energy, India.
\end{acknowledgement}

\end{document}